\documentclass[preprint]{elsart}
\usepackage[dvips]{graphicx}
\usepackage{epsfig}

\begin{document}
\begin{frontmatter}

\title{A Spin-Polaron Technique Utilized on Triangular-Lattice Antiferromagnet}
\author{Z.H. Dong}

\address{Department of Physics, Shanghai Jiao Tong University, Shanghai 200240, P. R. China}

\begin{abstract}

By expressing the Holstein-Primakoff transformation in a symmetric
form a modified spin-polaron technique utilized on
triangular-lattice antiferromagnet is developed. With the
technique, we have treated an extended $t$-$J$ model, calculated
the quasiparticle dispersion, and we also compared the the
dispersion with that obtained by other method.



\end{abstract}

\end{frontmatter}

\maketitle


\section{Introduction}

It is found that the cobalt oxide Na$_x$CoO$_2\cdot y $H$_2$O
($x\sim 0.35, y\sim 1.3$) has a triangular lattice in the CoO$_2$
planes\cite{Takada-Nature-2003,2006prl,prl-92,prl-92-3}. This
material is a fully frustrated system when only the
nearest-neighbor (NN) correlation is taken into account. So it
should be necessary to pay more attention on the
triangular-lattice antiferromagnet (TAFM) system. With this
motivation, we developed a modified spin-polaron technique to
discuss the quasiparticle dispersion of the TAFM.

\section{Holstein-Primakoff transformation}

In order to develop a spin-polaron technique on the TAFM, we first
express the Holstein-Primakoff (HP) transformation in a symmetric
form on the square- and triangular-lattice, respectively.

\subsection{Square lattice}
A square-lattice AFM consists two sublattices, one spin-up and the
other spin-down. We introduce a two-component vector
\begin{eqnarray}
\beta_i=\frac 1{\sqrt{2S}} \left(\matrix{\sqrt{2S-a_i^\dagger a_i}
\cr  a_i }\right),
\end{eqnarray}
(where $a_i$ being boson operators). Then, the HP transformation
can be expressed in terms of the vector $\beta_i$ as
\begin{eqnarray}
 S^z_i=&&S-a_i^\dagger a_i=S\beta_i^\dagger {\bf \sigma}_z \beta_i=S \beta_i^\dagger \pmatrix{1&0\cr 0&-1 \cr}
 \beta_i,\\
S^x_i=&&\frac 12 \left(a_i^\dagger \sqrt{2S-a_i^\dagger
a_i}+\sqrt{2S-a_i^\dagger a_i}a_i \right)= S\beta_i^\dagger {\bf
\sigma}_x \beta_i,\\
S^y_i=&&\frac i2 \left(a_i^\dagger \sqrt{2S-a_i^\dagger
a_i}-\sqrt{2S-a_i^\dagger a_i}a_i \right)= S\beta_i^\dagger {\bf
\sigma}_y \beta_i,
\end{eqnarray}
or
\begin{eqnarray}
\vec{s}_i=S \beta_i^\dagger \vec{\sigma}  \beta_i,
\end{eqnarray}
where $\vec \sigma$ is Pauli matrix.
\par
In spin-wave theory (SWT), in order to introduce only one type
Boson, a canonical transformation is usually performed to change
the N\'{e}el configuration
$|\uparrow\downarrow\uparrow\downarrow...>$ into a ferromagnetic
state with all spins up, {\em i.e.}, the $z$ axis of spin-down
sublattice must be upturned, forming the new local coordinate
$o-x'y'z'$. Now we investigate how the vector $\beta_i$ is rotated
with the coordinate rotation. Suppose the new coordinate is
obtained by rotating the old one by $180^\circ$ about its $x$
axis, with $z'$ pointing along the local N\'{e}el direction, the
direction of $x'$-axis is invariable and $y'$-axis is pointing
along $-y$. Accordingly, the spin components become as
\begin{eqnarray}
 \pmatrix{S_j^{\prime x}\cr S_j^{\prime y}\cr  S^{\prime z}_j}=\pmatrix {S_j^x \cr -S_j^y \cr -S^z_j }=R\pmatrix{S_j^x\cr
 S_j^y\cr S^z_j},
\end{eqnarray}
where
\begin{eqnarray}
 R=\pmatrix{1&0&0\cr 0&-1&0\cr 0&0&-1}
\end{eqnarray}
is SO(3) matrix. ($S^x_j,S^y_j,S^z_j$) are spin components in the
old coordinate, and ($S^{\prime x}_j,S^{\prime y}_j,S^{\prime
z}_j$) in the new local coordinate.
\par
With the coordinate rotation, $\beta_j$ become $\beta'_j$. We
suppose that 1) they are related through a indeterminate matrix
$u(R)$:
\begin{eqnarray}
\beta'_j=u(R)\beta_j,
\end{eqnarray}
 and 2) the HP transformation is unchanged in its form, {\em i. e.},
\begin{eqnarray}
\vec{s'}_j=S \beta_j^{\prime \dagger} \vec{\sigma} \beta_j^\prime.
\end{eqnarray}
Then, we have immediately the relation
\begin{eqnarray}
 \pmatrix {S'^x_j \cr S'^y_j \cr S'^z_j} =\pmatrix {S\beta_j^\dagger u^\dagger(R) \sigma_x
 u(R)\beta_j \cr S\beta_j^\dagger u^\dagger(R) \sigma_y u(R)\beta_j
 \cr S\beta_j^\dagger u^\dagger(R) \sigma_z u(R) \beta_j}.
\end{eqnarray}
From this equation the indeterminate matrix $u(R)$ can be easily
solved, and the results is
\begin{eqnarray}
 u(R)=\pmatrix{0&1\cr 1&0}.
\end{eqnarray}
Because the new coordinate is fixed on the spin-down sublattice
and the old one on the spin-up sublattice, the vector $\beta_i$
has the form of Eq. (1) on spin-up sublattice, and the form
\begin{eqnarray}
\beta_j^\prime=u(R)\beta_j=\frac 1{\sqrt{2S}} \left(\matrix{a_j
\cr \sqrt{2S-a_j^\dagger a_j} }\right)
\end{eqnarray}
on spin-down sublattice. If the prime is omitted and the spin-up
and -down sublattices are distinguished by indices, we have

\begin{eqnarray}
\beta_i=\left\{ \matrix{{\frac 1{\sqrt{2S}}
\left(\matrix{\sqrt{2S-a_i^\dagger a_i} \cr  a_i }\right)} &&
&(i\in {\rm spin-up\ \ sublattice}),\cr {\frac 1{\sqrt{2S}}
\left(\matrix{a_i \cr \sqrt{2S-a_i^\dagger a_i} }\right)}&& &(i\in
{\rm spin-down \ \ sublattice}).} \right.
\end{eqnarray}
The HP transformation can be merged into an unison form on the
both sublattices:
\begin{eqnarray}
\vec{s}_\alpha=S \beta_\alpha^\dagger \vec{\sigma}  \beta_\alpha,
\end{eqnarray}
with $\alpha=i,j$ corresponding to spin-up and -down sublattices,
respectively. It is easily verified that on both sublattices the
two-component vector satisfies the normal condition
\begin{eqnarray}
\beta_i^\dagger \beta_i=1.
\end{eqnarray}

\subsection{Triangular lattice}
Analog on the square-lattice AFM, now we express the HP
transformation on the TAFM. Unlike the square-lattice AFM, the
TAFM has three sublattices(called A, B and C) with three
$120^\circ$-N\'{e}el states, and their local coordinates can't be
simply divided into spin-up and spin-down sublattices, but into
three.
\par
Following Miyake \cite{doc21,doc22}, we define the local
(spatially varying) coordinates o-$x'y'z'$, with $y'$ pointing
along the old $z$ direction and $z'$ pointing along the local
$120^\circ$-N\'{e}el direction. When $x'$ is rotated by $0^\circ$,
$120^\circ$ and $240^\circ$ about $y'$ ($z$) axis respectively,
three new local coordinates are formed, which are fixed on the
sublattices A, B and C, respectively. In the three new coordinates
a spin operator has three forms:
\begin{eqnarray}
( S_i^{\prime x}, S_i^{\prime y}, S_i^{\prime z})=\left\{
\matrix{( S_i^y,S_i^{z},S_i^x )  &(i\in A) \cr
 ( -\frac
{\sqrt{3}}2 S_i^x-\frac 12 S_i^y, S_i^{z},-\frac 12 S_i^x+\frac
{\sqrt{3}}2 S_i^y ) &(i\in B)\cr  \left(\frac {\sqrt{3}}2
S_i^x-\frac 12 S_i^y ,S_i^{z},-\frac 12 S_i^x-\frac {\sqrt{3}}2
S_i^y \right) &(i\in
 C).} \right.
\end{eqnarray}
 We merge the three form into one
\begin{eqnarray}
 \pmatrix{S_i^{\prime x}\cr S_i^{\prime y}\cr S_i^{\prime
 z}}=R_{\alpha}^{-1} \pmatrix{S_i^ x\cr S_i^y\cr S_i^z} \;\;
 (\alpha=A,B,C),
\end{eqnarray}
Then the matrix $R_{\alpha}^{-1}$ can be easily resolved from the
Eqs.(16), and the inverse matrices are
\begin{eqnarray}
R_A= \pmatrix{0 & 0 & 1\cr 1& 0& 0 \cr 0& 1& 0},
 R_B=
\pmatrix{-\frac{\sqrt 3} 2& 0 & -\frac 12\cr -\frac 12& 0&
\frac{\sqrt 3} 2 \cr 0& 1& 0},
 R_C= \pmatrix{\frac{\sqrt 3} 2& 0&
-\frac 12\cr -\frac 12& 0& -\frac{\sqrt 3} 2 \cr 0& 1& 0}
\end{eqnarray}

Similarly in the Section 2.2, we here also introduce  a
two-component vector
\begin{eqnarray}
\beta_i(0)=\frac 1{\sqrt{2S}} \left(\matrix{\sqrt{2S-a_i^\dagger
a_i} \cr  a_i }\right)
\end{eqnarray}
in the old coordinate, and the HP transformation is still
expressed in terms of $\beta_i(0)$ as
\begin{eqnarray}
\vec{s}_i=S \beta_i^\dagger (0) \vec{\sigma}  \beta_i (0).
\end{eqnarray}
\par
When the coordinate is rotated, spin operator changes from
$\vec{s_i}$ to $\vec{s'_i}$, and the introduced matrix from
$\beta_i(0)$ to $\beta_i(\alpha)$ (where $\alpha=A, B$ $C$
corresponding to the three new coordinates, respectively). We
suppose the HP transformations on the new coordinates are
expressed in an unison form
\begin{eqnarray}
\vec{s}_i'=S \beta_i^\dagger (\alpha)\vec{\sigma} \beta_i(\alpha),
\end{eqnarray}
And we suppose also that
\begin{eqnarray}
\beta_i(\alpha)=u(R_\alpha)\beta_i(0).
\end{eqnarray}
From the Eqs.(21) and (22), we can express $\vec{s'}_i$ in terms
of $\beta_i(0)$:
\begin{eqnarray}
\vec{s'}_{i\alpha}=S \beta_i^\dagger(0) u^{\dagger}(R_{\alpha})
\vec{\sigma} u(R_{\alpha})\beta_i(0) \;\;\; (\alpha \in A, B, C).
\end{eqnarray}
The indeterminate matrix $u(R_{\alpha})$ can be determined by
substituting the Eqs. (20) and (23) into (17), and the results is
\begin{eqnarray}
u(R_{\alpha})=\pmatrix{\cos \alpha &-\sin \alpha \cr \sin \alpha
&\cos \alpha}
\end{eqnarray}
with $\alpha=0, 2\pi/3, -2\pi/3$ on sublattices A, B and C,
respectively. Eventually, Eq. (21) is just the HP transformation
in the three local coordinates of TAFM.

\section{Modified spin-polaron technique}

After expressing the HP transformation in terms of the introduced
matrix, we now develop a modified spin-polaron technique. The
spin-polaron picture was proposed early  by Schmitt-Rink Varma and
Ruckenstein \cite{dzh-11} to deal with the $t$-$J$ model on square
lattice\cite{dzh-14,dzh-15,dzh-17,dzh-18}. In this picture the
electron-annihilation operators are expressed as pure hole
operators or composite operators, for example,
\begin{eqnarray}
C_{i \downarrow}=h_i^\dagger s_i^\dagger
\end{eqnarray}
with $s_i^\dagger$ being the hard-core Bose operators. A similar
spin-polaron picture was proposed by Liu and Manousakis
\cite{dzh-14} by introducing two types of holes and two types of
spinons on spin-up and spin-down sublattices, respectively.
\par
Since the electronic operators $C_{i \sigma}$
($C_{i\sigma}^\dagger$)
 appear always in pairs in physical quantities (for example, the kinetic
operator $\sum_{<ij>,\sigma} C_{i\sigma}^\dagger C_{j \sigma}$,
current operator $\sum_{<ij>,\sigma}\vec{R}_i C_{i\sigma}^\dagger
C_{j \sigma}$, the Hamiltonian $H$, and for the $t$-$J$ model, the
single-occupancy constraint  $\sum_\sigma C_{i\sigma}^\dagger C_{i
\sigma}\le 1$), we should deal directly with the pair operator
$\sum_\sigma C_{i\sigma}^\dagger C_{j \sigma}=
C_{i\uparrow}^\dagger C_{j\uparrow}+ C_{i\downarrow}^\dagger
C_{j\downarrow}$, rather than the single electronic operators
$C_{i \sigma}$ ($C_{i\sigma}^\dagger$).
\par
Because the electron hopping operators $C_{i\uparrow}^\dagger C_{j
\uparrow}$ and $C_{i\downarrow}^\dagger
 C_{j \downarrow}$ correspond to the same hole hoppings from the site $i$
to $j$, the term $ C_{i\uparrow}^\dagger C_{j\uparrow}+
C_{i\downarrow}^\dagger C_{j\downarrow}$ should be proportional to
the hole hopping operators $h_i h_j^\dagger$, or
\begin{eqnarray}
C_{i\uparrow}^\dagger C_{j\uparrow}+ C_{i\downarrow}^\dagger
C_{j\downarrow}=\kappa_{ij} h_i h_j^\dagger.
\end{eqnarray}
The factor $\kappa_{ij}$ should be related to boson operators
$a_{i(j)}$ and $a_{i(j)}^\dagger$, and one may expand it in terms
of a series of these boson operators,
\begin{eqnarray}
\kappa_{ij}=A_0+A_1(a_i^\dagger+a_j)+A_2(a_i^\dagger
a_j+a_j^\dagger a_i)+...,
\end{eqnarray}
where $A_0,A_1,...$ are indeterminate coefficients. Determination
of them is determination of the modified spin-polaron technique.
\par
On the one hand, in terms of the electron operators and the Pauli
matrices, the spin operators can be expressed as ${\bf S}_i=\frac
12 \sum_{\alpha \alpha^\prime} C_{i\alpha}^\dagger {\bf
\sigma}_{\alpha\alpha^\prime}
   C_{i \alpha^\prime}$, and the corresponding $z$-component reads
\begin{eqnarray}
S_i^z=\frac 12 \sum_{\alpha \alpha^\prime} C_{i\alpha}^\dagger
{\bf \sigma}^z_{\alpha \alpha^\prime}
   C_{i \alpha^\prime}=S( C_{i\uparrow}^\dagger C_{i\uparrow}-C_{i\downarrow}^\dagger C_{i\downarrow}).
\end{eqnarray}
On the other hand, the component $s_z$ can be expressed as
\begin{eqnarray}
S_i^z
 =\beta_i^\dagger \pmatrix{1&0\cr 0&-1 \cr} \beta_i, \nonumber
\end{eqnarray}
So we have the relation
\begin{eqnarray}
 ( C_{i\uparrow}^\dagger C_{i\uparrow}-C_{i\downarrow}^\dagger C_{i\downarrow})
 =\beta_i^\dagger \pmatrix{1&0\cr 0&-1 \cr} \beta_i. \nonumber
\end{eqnarray}
If the negative sign is changed as positive, one immediately has
\begin{eqnarray}
 ( C_{i\uparrow}^\dagger C_{i\uparrow}+C_{i\downarrow}^\dagger C_{i\downarrow})
 &=&\beta_i^\dagger \pmatrix{1&0\cr 0&+1 \cr} \beta_i.
\end{eqnarray}
This is exactly true as it is an identity. Enlightened by this
relation, we may extend it from the same site to different site:
\begin{eqnarray}
( C_{i\uparrow}^\dagger C_{j\uparrow}+C_{i\downarrow}^\dagger
C_{j\downarrow})\propto \beta_i^\dagger \pmatrix{1&0\cr 0&+1 \cr}
\beta_j=\beta_i^\dagger \beta_j.
\end{eqnarray}
This extension implies that the factor $\kappa_{ij}$ have been
selected as
\begin{eqnarray}
\kappa_{ij}=\beta_i^\dagger \beta_j=\frac 1{\sqrt {2S}}
[(a_i^\dagger +a_j)-\frac 1{4S} (a_i^\dagger a_i a_j +a_i^\dagger
a_j^\dagger a_j)+...],
\end{eqnarray}
and the coefficients as $A_0=0, A_1=\frac 1{\sqrt{2S}}, A_2=0,
A_3=\cdot\cdot\cdot$. Finally, the Eqs. (26) and (31) make up the
modified spin-polaron transformation.
\par
It should be stressed that there may be other selections to $A$'s.
For example, one may suppose
$\kappa_{ij}=f(\beta_i,\beta^{\dagger}_i,\beta_j,\beta^{\dagger}_j)$
as long as the operators $\kappa_{ij}$ satisfy the necessary
requirements such as conjugation for the permutation of $i$ and
$j$, and unitarity when $i=j$. Different selection may correspond
to different magnon-holon coupling strength.

\par
Now we rewrite the modified spin-polaron transformation in a
compact form
\begin{eqnarray}
\sum_\sigma {C_{i\sigma}^\dagger C_{j\sigma}}=h_i
\beta_{i}^\dagger (\alpha) h_j^\dagger \beta_{j}(\beta),
\end{eqnarray}
where the index $\alpha$ $(\beta)$ is for distinguishing different
sublattices with the site $i$ ($j$) belonging to the sublattice
$\alpha$ ($\beta$).
\par
Because $\beta_{i}(\alpha)$ satisfies the normal condition
\begin{eqnarray}
\beta_{i}^\dagger(\alpha) \beta_{i}(\alpha)=1,
\end{eqnarray}
on the same site, with the modified spin-polaron technique the
no-double occupancy constraint is automatically built in:
\begin{eqnarray}
\sum_\sigma {C_{i\sigma}^\dagger C_{i\sigma}}=h_i
\beta_{i}^\dagger(\alpha) h_i^\dagger \beta_{i}(\alpha)=h_i
h_i^\dagger \leq1.
\end{eqnarray}

\section{Application of the spin-polaron technique on TAFM}

Now we use the modified spin-polaron technique to treat the TAFM.
Here we use the extended $t$-$J$ model to describe its physics.
Then, when the long-range correlations are taken into account, the
Hamiltonian reads
\begin{eqnarray}
&H =& H_{tt'}+H_J, \nonumber\\
& H_{tt'}=& -t\sum_{\langle ij \rangle_1\sigma}
{C_{i\sigma}^\dagger
 C_{j\sigma}}-t^\prime\sum_{\langle ij \rangle_2 \sigma}{C_{i\sigma}^\dagger
 C_{j\sigma}}-\mu\sum_i{C_{i\sigma}^\dagger
 C_{i\sigma}},\\
 &H_J =& J\sum_{\langle ij \rangle_1} {{\bf S}_i \cdot {\bf S}_j},
\end{eqnarray}
where the summations $\langle i,j \rangle_1$ and $\langle i,j
\rangle_2$ run over the NN and next-nearest neighbor (NNN) pairs
respectively and the operators $C_{i \sigma}^\dagger$ are
subjected to the single-occupancy constraint.
\par
The spin-spin correlation part $H_J$ of the Hamiltonian can be
treated with the HP transformation. In $k$ space the free part of
the spinon energy is
\begin{eqnarray}
H_J=\sum_k {\omega_k \alpha_k^\dagger \alpha_k},
\end{eqnarray}
where $\alpha_k$ are spinon operators. The spin-wave dispersion is
\begin{eqnarray}
\omega_k=\frac 12
JSz\sqrt{[(1+2\gamma_k^{(1)})(1-\gamma_k^{(1)})]}
\end{eqnarray}
where $ \gamma_k^{(1)}=\frac
1z\sum_{\vec{\delta}^{(1)}}{e^{i\vec{k}\cdot\vec{\delta}^{(1)}}} $
is the summations over the NN  sites. And the vectors
$+\vec{\delta^{(1)}}$ covers the six NN neighbors
$\vec{e}_x,-\vec{e}_x, -\frac 12 \vec{e}_x +\frac
{\sqrt{3}}2\vec{e}_y, \frac 12 \vec{e}_x -\frac
{\sqrt{3}}2\vec{e}_y, -\frac 12 \vec{e}_x -\frac
{\sqrt{3}}2\vec{e}_y$ and $\frac 12 \vec{e}_x +\frac
{\sqrt{3}}2\vec{e}_y )$, $\vec{e}_x$ being one of the basis
vectors, and $\vec{e}_y$ normal to $\vec{e}_x$. Eq. (38) is
exactly the same as that obtained by Leung and Runge \cite{doc22}.
\par
With the transformation Eq. (32) the Hamiltonian $H_{tt}$ can be
expressed by boson and hopping operators. If we preserve the
second order of bosons, the it reads
\begin{eqnarray}
&H_{tt'}=&H_t+H_{t'}, \nonumber\\
 &H_t\approx &  \frac 12 t\sum_{\langle ij
\rangle_1}h_ih_j^\dagger
  - \sqrt{\frac 3 {4S}} t \left[ \sum_{\langle
ij \rangle_1, j\in B} h_i h_j^\dagger(a_i^\dagger-a_j)-
\sum_{\langle ij \rangle_1, j\in C} h_i
h_j^\dagger(a_i^\dagger-a_j)\right] \nonumber \\
 & & - \frac 1{8S}
t\sum_{\langle ij \rangle_1}h_ih_j^\dagger  (a_i^\dagger a_i +
a_j^\dagger a_j-2a_i^\dagger a_j)
-\mu\sum_i{h_ih_i^\dagger} + {\rm H.c,}\\
&H_t'\approx & -t'\sum_{\langle ij \rangle_2}h_ih_j^\dagger \left[
1 - \frac 1{4S} (a_i^\dagger a_i + a_j^\dagger a_j-2a_i^\dagger
a_j) \right]+ {\rm H.c.}.
\end{eqnarray}
\par
In $k$ space with Bogliubov transformation, we have
\begin{eqnarray}
H_{tt} =\sum_k {\epsilon_k h_k^\dagger h_k}+H',
\end{eqnarray}
where $h_k$ are holon operators. The first term describes the
holon hopping, and holon dispersion is
\begin{eqnarray}
\epsilon_k=-\frac 12 [t\gamma_k^{(1)} -2t'\gamma_k^{(2)}],
\end{eqnarray}
where $ \gamma_k^{(2)}=\frac
1z\sum_{\vec{\delta}^{(2)}}{e^{i\vec{k}\cdot\vec{\delta}^{(2)}}} $
is the summations over the(NNN) sites. In Eq. (41) the second term
$H'$ describes the interaction between the holons and spinons.
\begin{eqnarray}
H' &=& \sum_{kp}{(V_{kp}^\dagger  h_k
h_p^\dagger\alpha_{k-p}^\dagger +V_{kp} h_p
h_k^\dagger\alpha_{k-p})},
\end{eqnarray}
where $V_{kp}$ is the coherence factors. Here we will not discuss
it in detail, but pay attention mainly on the holon dispersion.
\par
Eq. (42) gives out the holon dispersion when both NN and NNN
hoppings are included. If the NNN hopping is ignored, the spectrum
reduces to
\begin{eqnarray}
 \epsilon_k=-\frac 12 t\gamma_k^{(1)}.
 \end{eqnarray}
\par
It is a periodical function. Its amplitude is one half of
Trumper's \cite{add1} and only one sixth of Azzouz's \cite{add2}.
This means that the present dispersion is the least. Why? We know
that the TAFM is fully frustrated, and the ground state is very
disordered. The disorder certainly flattens the dispersion. So the
property of spin frustration is more fully maintained within the
present theory.
\par
In summary, after introducing a two-components matrix we express
the HP transformation in a symmetric form, based on this we
developed a modified spin-polaron technique. With the technique we
calculated the quasiparticle dispersion of an extended $t$-$J$
model. The dispersion is more reasonable than that obtained by
other methods. The present theory can fully describe the
frustrated TAFM.






\end{document}